\documentclass[letterpaper]{JHEP}
\usepackage[dvips]{graphicx}
\input epsf.tex
\def\be{\begin{equation}}
\def\ee{\end{equation}}
\def\bea{\begin{eqnarray}}
\def\eea{\end{eqnarray}}
\def\ba{\begin{array}}
\def\ea{\end{array}}

\def\part{\partial}

\def\cH{\cal H}

\def\hi{\hat \imath}

\preprint{SUGP-01/4-1\\
 hep-th/0105298}

\title{Polarization of the D0 ground state in Quantum Mechanics and Supergravity}
\author{Donald Marolf\thanks{E-mail address: {\tt marolf@physics.syr.edu}}
       and Pedro J. Silva\thanks{E-mail address: {\tt psilva@physics.syr.edu}}
       \\Physics Department, Syracuse University, Syracuse, New York 13244}
\date{DTP-yy-nn}
\abstract{The presence of a distant D4-brane is used to further investigate
the duality between M-theory and D0-brane quantum mechanics.
Although the D4-brane background fields are not strong enough to
induce a classical dielectric effect in the D0 system, a polarization
of the quantum mechanical ground state does result.
A similar deformation arises for
the bubble of normal space found near D0-branes
in classical supergravity solutions.
These deformations are compared and are shown to have the same structure in
each case.  Brief comments are included on the relation of D0-branes in this
background to D0-branes as instantons in the D4-brane field theory and an
appendix addresses certain infrared issues associated with 't Hooft scaling
in 0+1 dimensions.}
\begin{document}
%\pagestyle{}
%titel
%\rightline{IFT-UAM/CSIC-yy-n}
%\rightline{DTP-yy-nn}
%\rightline{SUGP-00/m-n}

%%%%%%%%%%%%%%%%%%%%%%%%%%%%%%%%%%%%%%%%%%%%%%%%%%%%%%%%%
\newpage
\section{Introduction}

In recent years the outlook within string theory has changed immensely.
While perturbative string calculations are still of interest, the new
cornerstones of the theory are non-perturbative duality conjectures.
Some of the most impressive such conjectures are those of
matrix theory \cite{bfss} and Maldacena's AdS/CFT conjecture
\cite{malda} and its generalizations
\cite{imsy}.   Predictions of these conjectures
are verified in ever increasing detail, including impressive recent
results \cite{KL} which receive no prediction from supersymmetry.

These conjectures relate the physics of certain gravitating systems to
that of specific non-gravitating gauge theories.
The dynamics of the dual field theories are deduced
from the low energy effective actions of the various non-abelian D-brane
systems (see e.g., \cite{Dft}).
The correspondences appear to rely on the
particular form of the non-abelian interactions.  Indeed, this form
can be traced to
important properties such as the large number of light degrees of freedom
that account for black hole entropy \cite{SV}.

Recently, several investigations \cite{tvr2,mye1}
have uncovered the form of certain non-abelian couplings of D-branes
to supergravity background fields.
It is natural to assume that the gauge theory/gravity
dualities continue to apply when couplings to such backgrounds are included.
Our goal here is to investigate this idea in a particular context by
studying the `polarization' of the Dp-brane bound state in the background
of a D(p+4)-brane.  For definiteness, we shall concentrate on the D0/D4
context.

In certain cases, the application of a Ramond-Ramond background field
to a D0-brane system induces a classical dielectric effect
and causes the D0-branes to deform into a non-commutative D2-brane
\cite{mye1}.  While the Ramond-Ramond fields of our D4-brane background will
not be strong enough to induce such a classical effect, they do modify
the potential that shapes the non-abelian character of the
quantum D0 bound state.  As a result, this bound
state is deformed, or polarized.

Two aspects of this
deformation will be studied and compared with the corresponding
supergravity system.  Fundamental to this comparison will be
the connection described by Polchinski
\cite{pol1} relating the size of the matrix
theory bound state to the size of the bubble of space that is well-described
by classical supergravity in the near D0-brane spacetime.
The near D0-brane spacetime is obtained by taking a limit in which
open strings decouple from closed strings and the result is a
ten-dimensional spacetime which has
small curvature and small string coupling when one is reasonably
close (though not too close) to the D0-branes.  However, if one moves
beyond some critical $r_c$, the curvature reaches the string scale.
As a result, the system beyond $r_c$
is not adequately described by the massless fields of classical supergravity.
Our goal is therefore to compare the deformations of the non-abelian D0-brane
bound state with the deformations of this bubble of `normal' space around
a large stack of D0-branes.

The deformations we study can be thought
of as induced by the presence of a background D4-brane.  Strictly speaking,
what we mean is that we consider an appropriate limit of the D0/D4 system
in which the open strings between D0's decouple from closed strings and
in which the strength of the D4-brane background is held constant.
While the details
are difficult to compute, certain scaling behaviors
will be deduced below.   As has become common
in string theory, we find that the quantum mechanical effects of the non-abelian
D0-brane couplings correctly reproduce the effects of classical supergravity in
the large $N$ limit.

The organization of the paper is straightforward.  We address the polarization
of the non-abelian ground state in section \ref{QM} and of the supergravity
bubble in section \ref{sugr}. We then close with some discussion in
section \ref{disc}, including some comments about other Dp/d(p+4) systems.
An appendix includes a more detailed
treatment of infrared effects and 't Hooft scaling in
0+1 dimensional Hamiltonian perturbation theory.  A consequence of this
analysis is that it strengthens the argument that Polchinski's
upper bound \cite{pol1} on the size of the D0 bound state in fact gives
the complete scaling with $N$.

\section{D0-brane Quantum Mechanics}
\label{QM}

We begin with the world-volume
effective field theory describing $N$ D0-Branes in the standard
D4-brane background. This action is a suitable
generalization of the action for a single D0-brane, consisting of
the Born-Infeld term together with appropriate Chern-Simon terms.
However, the full action encodes the Chan-Paton factors or non-abelian
degrees of freedom that arise from strings stretching between the D0-branes.
After presenting this effective theory, we specialize to the case of the
D4-brane background and study the resulting deformations of the bound state.

\subsection{Preliminaries}

We will be using the couplings first derived in  \cite{tvr1}. In
order to see the relation to the abelian case, it is convenient to
display the bosonic parts of these couplings using the action
proposed by Myers \cite{mye1}.  Fermions can then be added through
an appropriate supersymmetrization. While this action contains the
relevant terms from \cite{tvr2} and coincides with Tseytlin's
proposal \cite{tse1} in flat space-time, it is known to require
corrections at sixth order in the non-abelian field strength of
the world-volume gauge field \cite{bai1}.  However, such high
orders of accuracy will not be required for our discussion.

The first part of the
non-abelian D0 effective action
is the Born-Infeld term
\bea
S_{BI}=-T_0 \int dt \, STr\left( e^{-\phi}
\sqrt{-\left( P\left[E_{ab}+E_{ai}(Q^{-1}-\delta)^{ij}E_{jb}
\right]\right) \, det(Q^i{}_j)} \right)
\label{eq:1}
\eea
with
\be
 E_{AB} = G_{AB}+ B_{AB}\  \qquad { \rm and}\qquad
Q^i{}_j\equiv\delta^i{}_j + i\lambda\,[\Phi^i,\Phi^k]\,E_{kj}\ .
\label{eq:2}
\ee
In writing (\ref{eq:1})
we have used a number of conventions taken from Myers \cite{mye1}:
\begin{itemize}

\item{}
Indices to be pulled-back to the worldline (see below)
have been labelled by $a$.  For other indices,
the symbol $A$ takes values in the full set of space-time coordinates while
$i$ labels only directions perpendicular to the center of mass world-line.

\item{}
The parameter
$\lambda$ is equal
to $l_s^2$.
While this convention differs
by a factor of $2\pi$ from that of Myers \cite{mye1},  it will greatly
simplify our presentation.

\item{}
The center of mass degrees of freedom decouple completely and are
not relevant for our discussion. The fields $\Phi^i$ thus take
values in the adjoint representation of SU(N). As a result, the
fields satisfy $Tr \Phi^i=0$ and form a non-abelian generalization
of the coordinates specifying the displacement of the branes from
the center of mass. These coordinates have been normalized to have
dimensions of $(length)^{-1}$ through multiplication by
$\lambda^{-1}$.
\end{itemize}

The rest of the action is given by the
non-abelian Chern-Simon terms. These involve the non-abelian
`pullback' $P$ of various covariant tensors to the world-volume of
the D0-brane e.g. $P(C_a^{1}) = C^1_a \frac{\partial x^a}{\partial t}=
 C^{1}_0 + \lambda  C^{1}_i \frac{\partial
\Phi^i}{\partial t}$, where we have used the static gauge $x^0=t,
x^i = \lambda \Phi^i$ for a coordinate $x$ with origin at the D0-brane
center of mass.
The symbol $STr$ will be used to denote a trace
over the $SU(N)$ index with a complete symmetrization over the
non-abelian objects in each term.  In this way, the Chern-Simons
terms may be compactly written
\bea S_{CS}=\mu_0\int dt
STr\left(P\left[e^{i\lambda\,\hi_\Phi \hi_\Phi} ( \sum
C^{(n)}\,e^B)\right]\right)\ .
\eea

The symbol $i_\Phi$ is a non-abelian generalization of the
interior product with the coordinates $\Phi^i$,
\bea
i_\Phi \left(\frac{1}{2}C_{AB}dX^AdX^A\right) =
\Phi^iC_{iB}dX^B.
\eea

A by now familiar property of this action is that it leads to the
dielectric effect (or Myers effect) whereby a constant electric Ramond-Ramond
4-form field strength changes the classical ground state into a non-abelian
solution known as the `fuzzy two-sphere.' This solution is expected to represent a
D2-brane made out of non-abelian D0-branes \cite{mye1}.

\subsection{D0-branes in the D4 background}

In this work wish to study the `polarization' of the D0-brane bound state for
the specific case of $N$ D0-branes living in the
space-time generated by a D4-brane.
Such a background is defined by the metric $G_{AB}$, the dilaton
$\phi$, and the Ramond-Ramond 6-form field strength $F_{A_1A_2A_3A_4A_5A_6}$:
\bea
\label{D4}
&&ds^2_4 = {\cal H}_4^{-1/2} \eta_{\mu\nu}dX^\mu dX^\nu +
{\cal H}_4^{1/2}\delta_{mn}dX^{m}dX^{n} \nonumber \\
&&e^{-2\phi} = {\cal H}_4^{1/2} \nonumber \\
&&F_{01234m} = \partial_m {\cal H}_4^{-1},
\eea
with all other independent components of the field strength vanishing.
Here the space-time coordinates described by the index
$A$ have been partitioned into directions parallel to the D4-brane
(which we will label with a Greek index $\mu$) and directions
perpendicular to the D4-brane (which we label with a Latin index
$m$).  The function ${\cal H}_4 = 1 + \left( \frac{r_4}{|X|} \right)^3$
is the usual harmonic function of the D4-brane solution with
 $|X|^{\,2}=\delta_{mn}X^mX^n$ and with $r_4 = (gN_4)^{1/3}  l_s$
being the constant that sets the length scale of the supergravity
solution.

To expand the Born-Infeld action in this background we first
evaluate the pull-back $\frac{\partial x^a}{\partial
t}(E_{ab}+E_{ai}(Q^{-1}-\delta)^{ij}E_{jb}) \frac{\partial
x^b}{\partial t}$.
Since $E_{it}=0$, the only term not involving derivatives of
$\Phi^i$ is $g_{tt} = \frac{\partial x^0}{\partial t} E_{tt}
\frac{\partial x^0}{\partial t}$.  For the terms that do involve
$\Phi^i$, the $E_{ab}$ term exactly cancels the term $-g_{ai}
\delta^{ij} g_{jb}$.  Thus, since $B_{AB}=0$ we have only
\bea
\left(- P\left[E+E(Q^{-1}-\delta)E\right] \right)^{1/2}&& =
\left(- g_{tt} -\lambda^2 \partial_t \Phi^i \partial_t \Phi^j
Q^{-1}_{ij} \right)^{1/2} \nonumber \\
= &&\sqrt{-g_{tt}} \left( 1 +
{\lambda^2 \over 2}(g_{tt})^{-1}\partial_t\Phi^i\partial_t\Phi^j
g_{ij} \;\;+ \right. \nonumber \\
&&\left. {-i\lambda^3\over 2}(g_{tt})^{-1}\partial_t\Phi^i\partial_t\Phi^j
[\Phi^k,\Phi^l]g_{ik}g_{jl} +O(\lambda^4) \right).
\eea

The factor $\sqrt{det(Q^i_j)}$ generates the well known
potential $[\Phi,\Phi]^2=[\Phi^i,\Phi^j][\Phi^k,\Phi^l] g_{li} g_{jk}$
as well as higher order terms; i.e.
\be
\sqrt{det(Q)}=\left(1+{\lambda^2 \over
4}[\Phi,\Phi]^2+ { i\lambda^3\over 3!}[\Phi,\Phi]^3+O(\lambda^4)\right).
\ee
Since the D4-brane background has $e^{-\phi} \sqrt{-g_{tt}} =1$, we find
\bea
S_{BI}=&&-T_0 \int dt \, STr\left\{ 1 +
\lambda^2\left({1\over {2g_{tt}}}\partial_t\Phi^i \partial_t \Phi^jg_{ij} +
{1\over 4}[\Phi,\Phi]^2\right) + \right. \nonumber \\
&&\left. -\lambda^3\left( {i\over {2{g_{tt}}}}\partial_t\Phi^i\partial_t\Phi^j
[\Phi^k,\Phi^l]g_{il}g_{jk}+{i\over 3!}[\Phi,\Phi]^3\right) + O(\lambda^4) \right\}.
\eea
Finally, the dependence of the fields $g_{AB}$ on the non-abelian scalars
$\Phi^i$ must be determined through the Taylor expansions
\be
g_{tt}=\sum_{n=0}^\infty\frac{\lambda^n}{n!}
\Phi^{i_1} ... \Phi^{i_n}
\partial_{i_1...i_n}g_{tt}\;\;,\;\; {\rm and} \;\;
g_{ij}=\sum_{n=0}^\infty\frac{\lambda^n}{n!}
\Phi^{k_1} ... \Phi^{k_n}
\partial_{k_1...k_n}g_{ij}.
\ee
This yields the following form for the Born-Infeld action, where from now
on the symbols $g_{ij}, \partial_k g_{ij}, g_{tt}, \partial_k g_{tt}, etc.$
refer to the values of the fields at the D0-brane center of mass.

\bea
S_{BI}&=&-T_0 \int dt \, STr\left\{ 1 +
\lambda^2\left({1\over {2g_{tt}}}\partial_t\Phi^i \partial_t \Phi^j g_{ij} +
{1\over 4}[\Phi,\Phi]^2\right)  \right.\hspace{4cm} \nonumber \\
&+& \left. \lambda^3\left( {1\over 2}\partial_t\Phi^i\partial_t\Phi^j\Phi^k\partial_k(g_{tt})^{-1}+
{1\over {2g_{tt}}}\partial_t\Phi^i\partial_t\Phi^j\Phi^k\partial_kg_{ij}+
{1\over 2}[\Phi,\Phi^i][\Phi^j,\Phi]\Phi^k\partial_k g_{ij}\right) \right.\;\;\;\;\;\; \nonumber \\
&-&\left. \lambda^3\left( {i\over {2g_{tt}}}\partial_t\Phi^i\partial_t\Phi^j
[\Phi^k,\Phi^l]g_{il}g_{jk} +{i\over 3!}[\Phi,\Phi]^3 \right) + O(\lambda^4) \right\}
\eea
The $STr$ prescription removes the final two terms, which have been
displayed separately
on the third line.  A careful study shows that the symmetrized trace in
fact removes any term with an odd number of commutators $[\Phi, \Phi]$.

The only non-zero Chern-Simons terms involve
the RR five form $C^{(5)}$.  Hence, in direct analogue with
Myers \cite{mye1} we find the dipole coupling
\be
S_{CS}=\mu_0\int d\tau
STr\left\{{\lambda^3\over 10}
\Phi^i\Phi^j\Phi^k\Phi^l\Phi^mF_{0ijklm} +O(\lambda^4) \right\}.
\ee
Now the above approximations for the Born-Infeld and Chern-Simons
actions are valid as long as
the suppressed terms are of no significance. In the case of the Chern-Simons
action the $O(\lambda^{3+k})$ term involves $4+k$ factors of $\Phi$
and $k$ derivatives of $F_{0ijklm}$.  Thus, the $O(\lambda^4)$ term can
be neglected for $ F\gg \lambda \Phi^i\partial F$, and similarly for the
higher terms. Since the D4-brane background (\ref{D4}) is weak when the
D4-brane is far away as we require, one expects $\lambda
\Phi^i \sim \ell_s$.   For such $\Phi$, the $O(\lambda^4)$ terms
are in fact small $|X| \sim r_4$ or greater. Similarly, this condition
allows us to discard $O(\lambda^4)$ terms in the Born-Infeld action.
Note that a study of the strong field effects on D0-branes near the D4-brane
would require an understanding of the full non-abelian action to
all orders.  It is for this reason that we consider only distant D4-branes.

Imposing the above condition and assuming that the commutators are small,
we arrive at the following (bosonic) effective action:
\bea
\label{SwF} S_{eff.} = -T_0\lambda^2 \int dt \, STr\left\{
{1\over {2 g_{tt}}}\partial_t\Phi\partial_t \Phi +
{1\over 4}[\Phi,\Phi]^2 + \right.\hspace{4cm}&& \nonumber \\
\left. + \lambda \left( {1\over 2}\partial_t\Phi^i\partial_t\Phi^j\Phi^k\partial_k(g_{tt})^{-1}+
{1\over {2 g_{tt}}}\partial_t\Phi^i\partial_t\Phi^j\Phi^k\partial_kg_{ij}\;+ \right.\right. \hspace{1cm}&&\nonumber \\
\left. \left.+{1\over 2}[\Phi,\Phi^i][\Phi^j,\Phi]\Phi^k\partial_k g_{ij}+ \frac{1}{10}\Phi^{i}\Phi^{j}\Phi^{k}\Phi^{l}\Phi^{m}F_{\tau
ijklm}\right) \right\}.&&
\eea
These couplings also appear in
\cite{tvr2,tvr1}, along with the appropriate Fermion terms.
The Fermion terms are rather long, and little insight is gained
by writing them explicitly here.  Drawing from
\cite{tvr2,tvr1}, the Fermion terms will be introduced
as needed below.

\subsection{Deformations of the bound state}
\label{def}

One might begin with a discussion of classical solutions corresponding to
the above
effective action.  However, aside from the trivial commutative solution, one
does not expect to find any static solutions\footnote{The
literature \cite{indios,gjs,clt} contains some examples of
classical non-commutative solutions in similar (but non-supersymmetric)
systems.}.
We quickly note that, as opposed to the situation in \cite{mye1}, the
BPS character of the commutative ground state forbids any
non-abelian classical
solutions from having lower energy.  Furthermore, any classical
dielectric effect in this context would amount to the formation of
a D4/anti-D4 pair out of the D0-branes.  A D4-brane by itself
would remain static in the D4-background, while an anti-D4 brane
would fall toward the background D4-brane.  One therefore expects
any bound state of D4 and anti-D4 branes to fall as well. Indeed,
while one can find a static non-abelian solution (albeit an
unstable one)  for
(\ref{SwF}), the corresponding values of $\Phi$ are larger than the domain
of validity of the expansion (\ref{SwF}).
One expects such a solution to be eliminated by a more
complete treatment.

Nevertheless, we may expect that the non-abelian couplings to the background
affect the quantum bound state by
altering the shape of the potential and thus
the ground state wavefunction. Let us calculate the size of the
ground state by considering the expectation value of the squared radius
operator $R^2 \equiv \lambda^2Tr(\Phi^2) = \lambda^2 Tr(\Phi^i \Phi^j g_{ij})$.
Note that by passing to an orthonormal frame one finds a full SO(9) spherical
symmetry in the $O(\lambda^2)$ terms in our action.  Thus, to $O(\lambda^2)$
the expectation value of $Tr (\Phi^i \Phi^i g_{ii})$ is independent of $i$
and $R^2$ is the radius of the corresponding sphere measured in terms
of string metric proper distance.

Here we give a simple argument for the behavior of $\langle R^2 \rangle$
based on the usual 't Hooft scaling
behavior.  There are, however, several subtleties that are pointed out below.
A more complete discussion in terms of Hamiltonian perturbation theory is
given in the appendix.

Our strategy is to treat the couplings to the D4-brane fields as perturbations
to the D0-brane action in flat empty spacetime.
Thus, we divide (\ref{SwF}) into an `unperturbed action' $S_0$ and a
perturbation $S_1$. Note that as we place the N D0-branes far
from the D4-branes, the Ramond-Ramond coupling term can be written in the form
\begin{equation}
\Phi^{\mu_1}\Phi^{\mu_2}\Phi^{\mu_3}\Phi^{\mu_4}\Phi^m
F_{0 \mu_1 \mu_2 \mu_3 \mu_4 m} =
\frac{f}{\lambda^{1/2}}\Phi^{\mu_1}\Phi^{\mu_2}\Phi^{\mu_3}\Phi^{\mu_4}
\epsilon_{\mu_1 \mu_2 \mu_3 \mu_4} \Phi^{m} \frac{X^m}{|X|}
\end{equation}
where $f
= 3 (\frac{r_4^3\lambda^{1/2}}{z_\perp^4})
\approx 3{\cal H}_4^{-2}(\frac{r_4^3\lambda^{1/2}}{z_\perp^4}) $
is a scalar
dimensionless measure of the field strength.
Here $z_\perp$
is the distance between the N D0-branes and the D4-brane, and
$\epsilon_{\mu_1 \mu_2 \mu_3 \mu_4}$ is the antisymmetric symbol on four
indices.  In what follows we
treat all effects of the D4-brane only to lowest order in $({\cal H}_4-1)$
and $f = 3 (\frac{r_4^3\lambda^{1/2}}{z_\perp^4})$, so that $f
\approx {\cal H}_4^{-2}f$. With this understanding, the other $O(\lambda^3)$
terms are also proportional to $f$.

It will be useful to express the dynamics in terms of
rescaled fields and a rescaled time coordinate:
\be
\tilde \Phi^i = \lambda^{1/2} {\cal H}_4^{1/12}
 (gN)^{-1/3} \Phi^i\;\;,\;\;
\tilde \Theta = \lambda^{3/4} {\cal
H}_4^{1/8}(gN)^{-1/2}\Theta\;\;,\;\; \tilde t = \lambda^{-1/2}
{\cal H}_4^{-1/3} (gN)^{1/3}t. \ee where we have included the
fermionic fields $\Theta$ for completeness. This yields the action
\cite{ch}
\begin{equation}
\label{FermAct}
S_0 = - N \int d\tilde t \,STr \left( - \frac{1}{2} \partial_{\tilde t} \tilde
\Phi \partial_{\tilde t} \tilde \Phi + \frac{1}{4} [\tilde \Phi, \tilde \Phi]^2
+ {1\over 2} \tilde \Theta \dot{\tilde \Theta}
-{1\over 2}\tilde \Theta \gamma^i [ \tilde \Phi_i, \tilde \Theta] \right)
\end{equation}
and the perturbation
\bea
S_1&&=- \left[ (gN)^{1/3}  {\cal H}_4^{-1/12} f \right]
N\int d\tilde t\,STr\left({1\over 10}\tilde\Phi^{i}\tilde\Phi^{j}\tilde\Phi^{k}\tilde\Phi^{l}\tilde\Phi^{m} \epsilon_{
ijklm}+{1\over 2}\partial_t\tilde\Phi^i\partial_t\tilde\Phi^j\tilde\Phi^k
\frac{ \partial_k(g_{tt})^{-1}}{f}+\right. \nonumber \\
&&\hspace{3cm}\left. +{1\over {2}}\partial_t\tilde\Phi^i\partial_t\tilde\Phi^j\tilde\Phi^k
\frac{ \partial_kg_{ij}}{f}+
{1\over 2}[\tilde\Phi,\tilde\Phi^i][\tilde\Phi^j,\tilde\Phi]\tilde\Phi^k
\frac{\partial_k g_{ij}}{f}+ \hbox{Fermions}\right),\nonumber \\
&&\equiv - [(gN)^{1/3} {\cal H}_4^{-1/12}  f]\,\tilde S_1 .
\eea
Note that in writing
$\tilde S_1$ we have extracted a factor of $f$ from $S_1$.  The advantage
of this form is that both $S_0$ and $\tilde S_1$ are
manifestly independent of $g$, $\lambda$, and
$f$ while they depend on $N$ only through the overall factor and the trace.
The dependence of $S_0$ and $\tilde S_1$
on ${\cal H}_4$ is only though contractions with $g_{ij}$.
These could be further eliminated by passing to an orthonormal frame, so
the dynamics of
scalar contractions such as $\Phi^i\Phi^j g_{ij}$ will be independent of
${\cal H}_4$.

While we will not need the explicit form of the Fermion terms in $\tilde S_1$,
we will use the fact that each Fermion term can be obtained from the bosonic
terms by replacing three bosons with two fermions and an odd number of $\gamma^i$
matrices\footnote{Essentially the number of such $\gamma^i$ matrices is three,
though sometimes pairs $\gamma^i \gamma^i$
appear contracted together and so form a matrix
proportional to the identity that is not explicitly displayed.}.
This important property may be checked from the results of \cite{tvr1} and is
related to the structure of the associated superfields.
Note that the same relation holds
between the purely bosonic term and the Fermion term in the
potential for $S_0$.

As a result of this structure, the
Fermion terms in $\tilde S_1$ again contain no explicit factors
of $g$,  $\lambda$, or $f$ and have the same minimal dependence
on ${\cal H}_4$ and
$N$ as the purely bosonic terms.   It also follows that $S_1$
is antisymmetric under a total inversion of space in which
the bosons $\Phi^i$ are mapped to $-\Phi^i$ and the Fermions
are rotated by $\pi$, effectively mapping the $\gamma^i$ matrices to
$-\gamma^i$.

Let us now consider
the case $f=0$ and the corresponding ground state $\langle R^2 \rangle_0$.
We will think of this as the limit of small $\frac{\ell_s}{z_\perp}$, so
that we preserve ${\cal H}_4 \neq 1$.
Note that we have
\begin{equation}
\label{unpert}
\langle R^2 \rangle_0 = (gN)^{2/3} {\cal H}_4^{-1/6} \lambda
\langle Tr \tilde \Phi^2 \rangle_0.
\end{equation}
The factor $\langle Tr \tilde \Phi^2 \rangle_0$ is manifestly independent
of $g$, ${\cal H}_4$, and $\lambda$,
and the form of $S_0$ is the usual one associated with 't Hooft scaling
for which $\langle Tr \tilde \Phi^2 \rangle_0$ is also independent of $N$
in the limit of large $N$ with $gN$ fixed. This reproduces the results
of \cite{pol1,bfss}:
\begin{equation}
\label{qmr0}
\sqrt{\langle R^2 \rangle_0} \sim (gN)^{1/3} {\cal H}_4^{-1/12} \lambda^{1/2},
\end{equation}
where the product of $(g {\cal H}_4^{-1/4})^{1/3}$ can be viewed as the
natural dependence on the local string coupling $ g_{local} \equiv g e^\phi
= g{\cal H}^{-1/4}_4$ of the D4-brane background.

Of course, the usual 't Hooft scaling argument is stated in terms
of field theoretic
perturbation theory and scattering states.  As a result, there
may be some subtlety in applying it to 0+1 systems.  From the field
theoretic viewpoint, such subtleties are associated with the infrared
divergences typical of 0+1 dimensions.  Nevertheless, we may take
courage from the fact that (\ref{qmr0}) agrees with the upper bound in
\cite{pol1}. We will comment further on this below
and the reader can find a more thorough treatment in the appendix.

Let us now turn to the perturbed system. Considering the ground
state expectation value as the low temperature
limit of a thermal expectation value gives a Euclidean path integral
for $\langle R^2 \rangle$.  We wish to expand the factor $e^{-S_1} =
e^{-\left((gN)^{1/3} {\cal H}_4^{-1/12}  f \right) \tilde S_1}$ as
$1 - (gN )^{1/3} {\cal H}_4^{-1/12}  f  \tilde S_1 + (gN)^{2/3}
{\cal H}_4^{-1/6} f^2 \tilde S_1^2
- ...$.  Note that the order zero term gives just
$\langle R^2 \rangle_0$, the expectation value in the unperturbed
ground state.  The contribution from the
first order term then vanishes because
$\tilde S_1$ is anti-symmetric under a total inversion of space while
$R^2$, $S_0$, and the integration measure are invariant.

Thus, the leading contribution is of second order
in $\tilde S_1$ and we have
\begin{equation}
\label{s1div}
\frac{\langle R^2 \rangle -  \langle R^2 \rangle_0}{\langle R^2 \rangle_0}
\approx
(gN)^{2/3} {\cal H}_4^{-1/6} f^2
\frac{\langle T \left[(Tr \tilde \Phi^2)
(\tilde S_1)^2\right] \rangle_0}{\langle Tr\tilde \Phi^2 \rangle_0},
\end{equation}
where the $T$ represents time ordering.
Again the factor
$\langle T \left[ (Tr \tilde \Phi^2) (\tilde S_1)^2\right] \rangle_0$
is explicitly
independent of $g$, $\lambda$, ${\cal H}_4$,
and $f$. Furthermore, when written in
terms of $\tilde \Phi^i$ and $\tilde \Theta^i$,  $\tilde S_1$ and
$S_0$ have the typical form associated with 't Hooft scaling for which
$\langle Tr \tilde \Phi^2 \rangle$ should
be independent of $N$ in a system with action $S_0 + f \tilde S_1$
in the limit of large $N$ with $gN$ fixed.
Thus, we may express our final result as
\begin{equation}
\label{qmres}
\frac{\langle R^2 \rangle -  \langle R^2 \rangle_0}{\langle R^2 \rangle_0}
\sim (gN)^{2/3}  {\cal H}_4^{-1/6} f^2.
\end{equation}

Here again one may ask about subtleties of the 't Hooft limit and infrared
divergences in 0+1 dimensions.  In particular, the factors of $\tilde S_1$
in (\ref{s1div}) contain an integration over $\tilde t$.
These integrals almost certainly
diverge due to the fact that correlations do not die off at large times
in 0+1 dimensional systems.  To address such  concerns,
a more complete argument in terms
of Hamiltonian perturbation theory is provided in the appendix.
This consists of
regulating the infrared divergence and working through the 't Hooft
power counting for our case.  If, in analogy with the usual
't Hooft argument at strong coupling, one takes as input that one should
be able to read off certain properties of the full expectation values
from an asymptotic expansion, then the arguments in the appendix
can be said to give a proof of (\ref{qmres}).

\section{The D0-D4 system in supergravity}
\label{sugr}

Having considered the quantum mechanical description of the non-abelian
D0-brane bound state,
we now wish to compute a corresponding effect in classical
supergravity. We seek a BPS solution containing both D0's and D4's
with the D0's being both fully localized and separated from the
D4-branes.  It is conceptually simplest to discuss the full D0/D4
solution and then take a suitable decoupling limit.
Such full solutions are known exactly, but only as an
infinite sum over Fourier modes \cite{mar1}. As a result, we find
it more profitable here to follow a perturbative method as suggested by the
quantum mechanical calculation above.  We therefore expand
the supergravity
solution in $f$, the magnitude of the Ramond-Ramond
4-form field strength at the location of the zero-branes.

\subsection{The Perturbed Solution}

Let us consider a BPS system of D4-branes and $N$
D0-branes with asymptotically flat boundary conditions. Using
the usual isotropic ansatz in the appropriate
gauge reduces the problem to solving the equations
\cite{HM,Tseytlin,BREJS,AIRV}

\bea
&&\left(\partial_{\bot}^2+{\cal H}_4\partial_{\|}^2\right)
      {\cal H}_0 = 0, \nonumber \\
&&{\cal H}_4=1+\left(\frac{r_4}{|X|}\right)^3, \label{eq:6}
\eea

\noindent
where as before the D4-brane lies at $X^m=0$,
$|X|^{\,2}=\delta_{mn}X^mX^n$, and
${\cal H}_4$ and ${\cal H}_0$ are the `harmonic' functions for the
D4-brane and D0-brane respectively.
The two relevant derivative operators are a flat-space Laplacian
($\partial_{\|}^2 \equiv \sum^{\mu=4}_{\mu=1}\partial_\mu\partial_\mu$)
associated with the directions parallel to the D4-brane and
another ($\partial_{\perp}^2 \equiv
\sum^{m=9}_{m=5}\partial_m\partial_m$) associated with the perpendicular
directions.

In order to treat the D4-branes as a perturbation, we place them
far away from the D0-branes. It is convenient to change to new
coordinates $x^m$ (lowercase) whose origin is located at the D0
singularity.  One of these coordinates is distinguished by running
along the line connecting the D0- and D4-branes.  Let us call this
coordinate $x_\perp$.  The other four $x^m$ coordinates will play
a much lesser role. Introducing the distance $z_{\perp}$ between
the D0- and  D4-branes and expanding ${\cal H}_4$ to first order
about the new origin yields \bea {\cal H}_4 \approx {\cal
H}_4(x=0) - 3\left( \frac{r_4}{z_\perp} \right)^3
\left(\frac{x_{\perp}}{z_\perp}\right) \equiv {\cal H}_4(0) +
\delta{\cal H}_4. \label{eq:cs} \eea Here we have used $z_\perp\gg
(r_4,x_\perp)$, since  the D4-branes are located far away. Note
that fixing $z_\perp$ sets the location of the D0 singularity
relative to the D4-brane.  This is much like the fixing of the D0
center of mass degrees of freedom in section 2 as both set the
overall location of the D0 branes.  However, as we will see, it is
not clear in general that the position of the singularity
corresponds precisely to the center of mass. Equation \ref{eq:6}
can be solved by expanding ${\cal H}_0$ in terms of
$\delta{\cH}_4\,$ i.e. $\,{\cH}_0={\cH}_{00}+{\cH}_{01}+... $
where ${\cal H}_{0n}=O(\delta {\cal H}_4^{\,\,\,n})$. We find \bea
\label{eqtos} &&(\partial^2_\perp + {\cal H}_4(0)
\partial_\parallel^2) {\cH}_{00}= 0, \;\; {\rm so \ that}
\;\;{\cH}_{00}=1
+\left(\frac{r_0}{r}\right)^7, \nonumber \\
&&(\partial^2_\perp + {\cal H}_4(0) \partial_\parallel^2) {\cH}_{01}=
\delta{\cH}_4\partial^2_\parallel{\cH}_{00} \;\;{\rm and} \nonumber \\
&&(\partial^2_\perp + {\cal H}_4(0) \partial_\parallel^2) {\cH}_{02}=
\delta{\cH}_4\partial^2_\parallel{\cH}_{01}.
\eea
Here we have introduced $r^2= |x|^2 \equiv \delta_{mn}x^m
x^n+ {\cal H}_4^{-1}(0)
\delta_{\mu \nu}x^\mu x^\nu$, a sort of coordinate distance from the D0-brane.
Note that this $r$ does in fact label spheres of symmetry for the
unperturbed solution ${\cal H}_{00}$.

Since the D4-brane background has altered the background metric for the
D0-brane system, this will change certain familiar normalizations.
We therefore note that total electric flux from the D0-branes
must equal the number $N$ of D0-brane charge quanta
$\frac{g \ell_s^7}{60\pi^3}.$
Since the D4-brane is far away, we may compute this flux in a
region where ${\cal H}_0 \approx 1$ but where ${\cal H}_4 = {\cal H}_4(0)$.
The result yields $\frac{gN\ell_s^7}{60 \pi^3} = {\cal H}_4^2 r_0^7.$

The above equations (\ref{eqtos})
are easily solved in terms of Green's functions.
To first order the solution is:
\be
\label{H1}
{\cH}_{01}= 7r_0^7\lambda^{-1/2}f \int {dy^9 \frac{1}{\mid
x-y\mid^7}\frac{y_\perp(9y^2_\parallel-y^2)}{y^{11}}},
\label{eq:3}
\ee
where $f= \frac{3\lambda^{1/2}r_4^3}{z_\perp^4}$.  Here we have
used ${\cal H}_4 \approx 1$ since (\ref{eq:3}) is already proportional
to $f$.

While the integral is difficult to compute explicitly, the important
features of ${\cal H}_{01}$ can be readily deduced.  Note for example that
under a rescaling of coordinates
$y,x \rightarrow \beta y, \beta x$ the function ${\cal H}_{01}$ scales
homogeneously as $\beta^{-6}$.  As a result, we may write
\be
{\cH}_{01} = \frac{\omega_1 f}{x^6} r_0^7 \lambda^{-1/2},
\ee
where $\omega_1$ is an unknown dimensionless function of the
angles associated with the direction cosines $x^A/r$.   Furthermore,
${\cal H}_{01}$ is even under any $x^\mu \rightarrow x^\mu$ and under
any $x^m \rightarrow x^m$ for $x^m \neq x_\perp$.  However, we see that
${\cal H}_{01}$ is odd under $x_\perp \rightarrow - x_\perp$.

It is worthwhile to pause and understand the meaning of this anti-symmetry.
One consequence is that first order effects of ${\cal H}_{01}$ on the
spacetime curvature at the original bubble wall for a point with $x_\perp > 0$
are clearly opposite to those at the corresponding point with $x_\perp < 0$.
Thus, if the curvature increases in one place it must decrease in the other.
Since the perturbed wall is  by definition the place where the metric
has structure at the string scale, we see that if the wall moves
toward the origin for $x_\perp > 0$ then it must move {\it away}
from the origin for $x_\perp <0$.  Thus, the ${\cal H}_{01}$ term provides
an (angle dependent) {\it shift} of the bubble so that it is no longer
centered on the D0-brane center of mass.  In particular, this has no
effect on the {\it size} of the bubble.

We will return to this point below,
but as a consequence we will need to study the second order term
${\cal H}_{02}$.  We find

\be
\label{H2}
{\cH}_{02}= 49r_0^7\lambda^{-1}f^2 \int {dy^9 \frac{\omega_1}{\mid
x-y\mid^7}\frac{y^2_\perp(9y_\parallel^2-y^2)}{y^{11}}},
\ee
Again,  scaling arguments show that we have
\be
{\cH}_{02} = \frac{\omega_2 f^2}{x^5},
\ee
where $\omega_2$ is again a dimensionless function of the angles.
This time, however, ${\cal H}_{02}$ is even under $x^A \rightarrow x^A$
for any $A$.  As a result, ${\cal H}_{02}$ directly encodes a change in
the size of the bubble.  Although we have not
evaluated the integrals (\ref{H1}), (\ref{H2}) explicitly, both expressions
converge and could be computed numerically.
\begin{figure}
\begin{center}
\includegraphics[angle=0,scale=0.6]{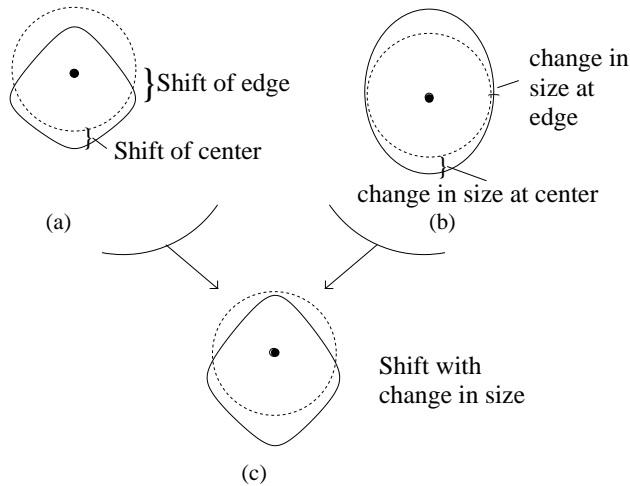}
\caption{{\it Two types of corrections to the radius of the N
D0-brane system, first the shift of the boundary and second
the change in size.  Note that both effects are angle dependent.
The unperturbed bubble wall has been shown as a dashed circle.}}
\label{fig:f2}
\end{center}
\end{figure}

\subsection{Analyzing the Deformations}

Let us now calculate the size of this solution. One might try to
measure the size of the D0-brane configuration by studying properties
of the Fourier transform (as in \cite{SM,mar1}) or by identifying where the
potential becomes of order one.  However, it is
clear that we wish to follow Polchinski \cite{pol1} and use the
measure that successfully reproduces the size of the unperturbed
D0-brane bound state.  While \cite{pol1} described this
correspondence in terms of quantities associated with the
eleven-dimensional metric,  we  prefer to use the ten-dimensional string
metric as this
has been the setting for our discussion so far.

This means that we should locate the surface
enclosing the D0-brane singularity where the string metric is so strongly
curved that it has structure on the string scale.  Inside
this surface is a bubble of space that is well described by classical
supergravity.  However, when $r$ is large
the proper distance $2\pi {\cal H}_4^{1/4} {\cal H}_0^{1/4} r$
around the sphere
enclosing the origin may still be on the order of $\ell_s$ so that
the metric clearly has structure on the string scale.  In particular,
one might think of strings encircling
the bubble itself.  The region inside this surface is to be
associated with the quantum D0-bound state, and one expects the bubble of
`normal' space and the bound state to have corresponding sizes and shapes.

Now, given the correspondence between the $r$ of supergravity
isotropic coordinates and the non-abelian $\Phi^2$ in D0 quantum mechanics
in the absence of the D4-brane, it is natural to expect a correspondence
between the
$R^2 = Tr (\Phi^i \Phi^j g_{ij})$ of section \ref{QM} and
the supergravity quantity $R^2 = {\cal H}_4^{1/2} \delta_{mn}x^m x^n
+ {\cal H}_4^{-1/2} \delta_{\mu \nu}x^\mu x^\nu = {\cal H}_4^{1/2} r^2$.
The edge of the bubble lies at the value of $R \equiv {\cal H}_4^{1/4} r$
for which $\ell_s \sim r {\cal H}_0^{1/4} {\cal H}_4^{1/4}
 = R {\cal H}_0^{1/4}(r).$  This yields the relation
\bea
\label{Reqn}
R \sim && \ell_s [{\cH}_0(r)]^{-1/4} \sim \nonumber \\
&& \ell_s {\cH}_{00}^{-1/4}\left(
1-\frac{{\cH}_{01}}{4{\cH}_{00}}-\frac{{\cH}_{02}}{4{\cH}_{00}}+
\frac{5}{20}\left(\frac{{\cH}_{01}}{4{\cH}_{00}}\right)^2 +
O(f^3)\;\right).
\eea

In order to make connections with the quantum mechanical
calculations, we wish to consider this system in the decoupling
limit $g \rightarrow 0$ with fixed $gN$ and $r/\ell_s \sim g^{1/3}
\rightarrow 0$.  Note that this scaling may seem more familiar when expressed in
terms of the eleven-dimensional plank mass $M_{11} = g^{-1/3} \ell_s^{-1}$
as it holds fixed the dimensionless quantity $rM_{11}$.
We see that the corresponding asymptotic behavior of ${\cal H}_{00}$ is
given by
\begin{equation}
{\cal H}_{00}  \approx \left( \frac{r_0}{r} \right)^7 = \frac{ gN \ell_s^7}
{{\cal H}_4^{1/4}  R^7}.
\end{equation}
As a result, keeping only the order zero term yields an unperturbed value
$R_0$ of $R$ given by
$R_0 \sim (gN)^{-1/4} \ell_s^{-3/4}
{\cal H}_4^{7/16}  R_0^{7/4}$, or
$R_0 \sim (gN)^{1/3} \ell_s
{\cal H}_4^{-1/12}$ in agreement with (\ref{unpert}).

Recall that the effect of the ${\cal H}_{01}$ term is to shift the bubble
by an angle-dependent amount but not to change the size of the bubble.  Due to
the angle-dependence, concepts like the radius $R$ of the bubble also become
angle-dependent.  However, it is the average $\langle R^2\rangle$
of $R^2$ over the bubble
that we expect to compare with expectation values in
the quantum mechanical ground state.    Taking such an average, it is
clear that the term in (\ref{Reqn}) that is linear in ${\cal H}_{01}$ has no
effect on $\langle R^2 \rangle$.  Of course, a shift of the bubble
away from the origin will contribute to $\langle R^2 \rangle$ at second
order, and this effect is governed by the term
$\frac{5}{20}\left(\frac{{\cH}_{01}}{4{\cH}_{00}}\right)^2$.
This is in agreement with our
quantum mechanical calculation where the effect on $\langle R^2 \rangle$
appeared only at second order in $f$.

Averaging (\ref{Reqn}) over the bubble removes the linear term in
${\cal H}_{01}$, but the second order terms continue to contribute.
Note that they both are of the form
\begin{equation}
\frac{{\cH}_{02}}{4{\cH}_{00}}
\sim
\frac{5}{20}\left(\frac{{\cH}_{01}}{4{\cH}_{00}} \right)^2 \sim
(const) f^2 \langle
R \rangle^2.
\end{equation}
After taking the decoupling
limit we find
\be
\label{afdc}
\langle R \rangle \sim {\langle R\rangle ^{7/4} \over (gN)^{1/4}
{\cal H}_4^{1/16}
\ell_s^{3/4}}\left(1 + (const)
f^2\langle R\rangle^2\right),
\ee
since to the necessary order of accuracy we have
$\langle R \rangle^n = \langle R^n \rangle$.
Solving for $\langle R \rangle$ to second order in $f$ yields
the relation
\be
\langle R \rangle \sim  (const_1) l_s(gN)^{1/3} {\cal H}_4^{-1/12}
+ (const_2) {1\over l_s^2 }
[l_s(gN)^{1/3}{\cal H}_4^{-1/12} ]^2l_s(gN)^{1/3}{\cal H}_4^{-1/12} f^2, \
\ee
or
\be
\label{cgres}
\frac{\langle R\rangle^2 - R_0^2}{R^2_0} \sim f^2{\cal H}_4^{-1/6} (gN)^{2/3},
\ee
in agreement with the quantum mechanical results of (\ref{qmres}).

Finally, we may go one step further and interpret (\ref{cgres}) in
terms of the (angle dependent) shift of the bubble away from the
D0 singularity\footnote{i.e., the shift of the center of the
bubble away from the D0 singularity.  Unfortunately, it is unclear
how to identify the separation of the center and the singularity
in terms of the gauge theory description of section 2.}
($shift$) and the change in the size
($\Delta size$), let us write $R = R_0 + shift + \Delta size$
where we think of each quantity as depending on the direction in
which it is measured from the D0-brane singularity and where we
take $shift$ to be proportional to $f$ and $\Delta size$ to be
proportional to $f^2$ as indicated by our discussion above.
Averaging over all directions we find
\begin{equation}
\langle (R_0 + shift + \Delta size)^2 \rangle - R_0^2 =
R_0 \langle shift \rangle + R_0 \langle  \Delta size \rangle
+ \langle (shift)^2 \rangle + O(f^3),
\end{equation}
where of course, $\langle shift \rangle =0$ to lowest order\footnote{The
contribution to $\langle shift \rangle$ from those points that are not moved
across the $x_\perp=0$ plane averages to zero, but the
contribution from those points that are moved across this plane does not.
However, the fraction of points moved across the plane
is proportional to $shift/R_0$, so this simply
generates an additional second order contribution of size $(shift)^2$.}.

Since both second order
contributions to (\ref{Reqn}) were of the same order, we have
\begin{eqnarray}
\label{sds}
\Delta size &\sim& R_0 f^2 (gN)^{2/3} {\cal H}_4^{-1/6}, \ \ {\rm and} , \cr
shift &\sim& R_0 f (gN)^{1/3} {\cal H}_4^{-1/12}.
\end{eqnarray}
This is more information than was obtained from the quantum
mechanical calculations as in that case we were unable to separate
the contributions of the shift and change in size.  Nevertheless,
(\ref{sds}) is consistent with the quantum mechanical results.

\section{Discussion}
\label{disc}

In the above work we have analyzed deformations of the non-abelian
D0-brane bound state and of the bubble of `normal space' in the near D0-brane
supergravity solution.  In each case, the deformations are polarizations
induced by the presence of a distant D4-brane and the deformations were shown
to have corresponding scaling properties.  This supports the idea that
the gravity/gauge theory duality associated with D0-branes can be
extended to include couplings to nontrivial backgrounds such as those
discussed in \cite{tvr1,tvr2,mye1}.  A part of this was the analysis  in
the appendix of infrared issues associated with 't Hooft scaling in 0+1
dimensions.  This in turn strengthens the argument that Polchinski's
upper bound \cite{pol1} on the size of the D0-brane bound state in fact
gives the full scaling with $N$.

One might expect our polarization effects
to be particularly interesting when the D0- and D4-branes are close together
and the effects are strong.  Indeed, it is known that the D0-brane
contribution to the supergravity solution is greatly distorted in the limit
of zero separation \cite{mar1}.  However, pursuing such an analysis in
D0-brane quantum mechanics would require an understanding of the full
non-abelian dynamics as a truncation to $O(\lambda^3)$ or any other
finite order would no longer be sufficient.

It is worth taking a moment to contrast the polarization effects
studied here with those studied in \cite{mar1} which were argued
to be dual to certain effects in the D4-brane field theory.  While
both works considered a change in the characteristic size of the D0-brane
supergravity solution induced by the D4-brane, the details are rather
different.  The previous work \cite{mar1} used a measure of the D0-brane
size in supergravity that followed from purely classical considerations.
In particular, the curvature did not reach the string scale near the edge
of the D0-brane region and the corresponding size could be expressed
entirely in terms of supergravity quantities without using any explicit
factors of $\ell_s$.   In contrast, in the present work we have followed
Polchinski \cite{pol1} and used a measure of size in which a supergravity
result (the distance around the D0-brane bubble) is explicitly set
equal to $\ell_s$.  It is therefore clear that these works use
quite different measures of the classical `size' of the D0-brane solution
and discuss different physics.

While this is clear with hindsight, it is also somewhat surprising.
Let us recall that the discussion of \cite{mar1} compared the D0-branes
in a supergravity solution with D0-branes as represented by solitons (Yang-Mills
instantons) in the D4-brane field theory.  The associated measure of
size in supergravity was argued to correspond to the scale size of the
instantons.  One might well think it natural that the scale size of
such instantons be connected to the size of a D0-bound state in matrix
theory.  We see, however, that the relation between matrix theory and
instantons in the D4-brane theory must be more subtle.  It would be
of interest to understand this in detail.

Finally, let us consider extensions to other Dp/D(p+4) systems.
Having set up the framework, such calculations are straightforward.
Adapting our covariant scaling argument to the p+1 dimensional
unperturbed bosonic action
\begin{equation}
S_0 = \frac{\lambda^2}{g} \int d^{p+1}x \ STr  \left[  e^{-\phi}
\left( \frac{1}{2}
( \partial \Phi)^2 + \frac{1}{4} [\Phi,\Phi]^2 \right) \right]
\end{equation}
suggests the rescaling
\be
\tilde \Phi^i = \lambda^{1/2}
 (gN{\cal H}_{p+4}^{-1/4})^{-1/(3-p)} \Phi^i\;\;,\;\;
\tilde x = \lambda^{-1/2}
(gN {\cal H}_{p+4}^{-1/4} )^{1/(3-p)}x,
\ee
where ${\cal H}_{p+4}$ is the D(p+4)-brane harmonic function associated
with the background in which the Dp-branes are to be placed.
This yields
\begin{equation}
\frac{\langle R^2 \rangle -  \langle R^2 \rangle_0}{\langle R^2 \rangle_0}
\sim (gN {\cal H}_4^{-1/4})^{2/(3-p)}  f_{p+4}^2,
\end{equation}
where $f_{p+4} = \ell_s \partial_{x_\perp} {\cal H}_{p+4}$ is again a
dimensionless measure of the field strength produced by the distant
D(p+4)-brane.

The supergravity calculation is correspondingly straightforward.  The only
change is to modify a single power in (\ref{afdc}) to yield
\be
\langle R \rangle \sim {\langle R\rangle^{(7-p)/4}
\over (gN{\cal H}_{p+4}^{1/4} )^{1/4}
\ell_s^{3/4}}\left(1 + (const)
f_{p+4}^2\langle R\rangle^2\right),
\ee
Solving this equation as before one finds
\be
\langle R \rangle \sim  (const_1) l_s(gN {\cal H}_{p+4}^{-1/4} )^{1/(3-p)}
+ (const_2) {1\over l_s^2 }
[l_s(gN{\cal H}_{p+4}^{-1/4} )^{1/(3-p)}]^2l_s(gN{\cal
H}_{p+4}^{-1/4})^{1/(3-p)} f_{p+4}^2, \
\ee
or
\be
\frac{\langle R\rangle^2 - R_0^2}{R^2_0} \sim f_{p+4}^2
(gN{\cal H}_{p+4}^{-1/4})^{2/(3-p)}.
\ee

Again, we find complete agreement between the quantum mechanical and
supergravity calculations.  However, for $p=3$ the above `solutions'
diverge and there is in fact no solution of this form.
Larger branes are troubled by the issues that the associated
quantum field theories are not renormalizable and that there is in fact
no regime in which the field of the $p+4$ brane is weak.
However, the results should be meaningful for the cases $p=-1$, $p=1$ and
$p=2$.  Regarding the issue of infrared divergences in the quantum mechanical
argument, these should not be important for $p=2$ and they are manifestly
absent for $p=-1$ since the worldvolume of an instanton is compact.
For $p=1$ one assumes that a more careful argument, perhaps along the lines
of that in the appendix, should be able to deal with any infrared
divergences.  Thus our conclusions should extend to these cases as well.

%%%%%%%%%%%%%%%%%%%%%%%%%%%%%%%%%%%%%%%%%%%%%

\acknowledgments

The authors would like to thank Mark Bowick, Amanda Peet, Joe
Polchinski, and Joel Rozowsky
for useful discussions.  This work was supported in part by
NSF grant PHY97-22362 to Syracuse University,
the Alfred P. Sloan foundation, and by funds from Syracuse
University.

%%%%%%%%%%%%%%%%%%%%%%%%%%%%%%%%
\vspace{1cm}
\noindent
{\bf \large Appendix}
\appendix
\vspace{.5cm}

\section{Hamiltonian treatment of the D0-brane quantum mechanics}

In this appendix we provide a more complete argument for the scaling with
$f$, $g$, $\lambda$, ${\cal H}_4$,
and $N$ of the deformations in the quantum mechanical
bound state.   The primary goal is to treat the infrared behavior with
due care.  This we succeed in doing, though certain technical assumptions
will need to be introduced as discussed below.  It is convenient to
separate the discussion of the dependence on $g$, $\lambda$, ${\cal H}_4$,
and $f$ (which
essentially translates the arguments of section 2 into Hamiltonian
language) from the discussion of the dependence on $N$.  The latter
will involve reconsidering the derivation of 't Hooft scaling  while
controlling the infrared behavior.  This strengthens the argument
that Polchinski's upper bound \cite{pol1} on the size of the D0 bound state
indeed gives the full scaling with $N$.

\subsection{Scaling with $g$, $\lambda$, ${\cal H}_4$, and $f$}

Since we work with a 0+1 system, the most control can
be obtained in the Hamiltonian formulation.   As in section \ref{def}, it is
useful to introduce rescaled bosonic fields
\be
\tilde \Phi^i = \lambda^{1/2} {\cal H}_4^{1/12} (gN)^{-1/3}\Phi^i \,\,, \,\,
\tilde P_i = \lambda^{-1/2} {\cal H}_4^{-1/12} (gN)^{1/3}P_i,
\ee
where $P_i$ is the (matrix valued) canonical momentum conjugate to
$\Phi^i$ and where $\tilde P_i$ is conjugate to
$\tilde \Phi^i$.  We also introduce the rescaled Fermions
$\tilde \Theta = \lambda^{3/4} {\cal H}_4^{1/8}
(gN)^{-1/2}\Theta$ and observe from the
normalization of the actions (\ref{SwF}) and (\ref{FermAct}) that  the rescaled
Fermions satisfy anti-commutation relations of the form $\{ \tilde
\Theta, \tilde \Theta \}_+ \sim \frac{1}{N}$.  In particular, these
anti-commutation relations are independent of $g$, $\lambda$,
${\cal H}_4$, and $f$.

In order to establish the scaling with $f$, $g$, ${\cal H}_4$, and $\lambda$,
it is useful to proceed much as in section 2 and
write the Hamiltonian $H$ associated with the
action (\ref{SwF}) in form $H_0 + H_1$ for
\bea
H_0 &&\equiv \left[{(gN)^{1/3}{\cal H}_4^{-1/12}
/\lambda^{-1/2}}\right]
\tilde H_0, \nonumber \\
&&=\left[{(gN)^{1/3}{\cal H}_4^{-1/12} / \lambda^{-1/2}}\right]
NSTr\left\{ \frac{1}{2}(N^{-1}\tilde P^i)(N^{-1} \tilde P_i)
+ \frac{1}{4}[\tilde \Phi,\tilde \Phi]^2
+ \hbox{Fermions} \right\}, \nonumber \\
H_1 &&\equiv \left[{(gN)^{2/3}{\cal H}_4^{-1/6} f / \lambda^{-1/2}}\right]
\tilde H_1, \nonumber \\
&&= \left[{(gN)^{2/3}{\cal H}_4^{-1/6} f / \lambda^{-1/2}}\right]
NSTr\left\{\frac{1}{10}\tilde \Phi^{a} \tilde
\Phi^{b} \tilde \Phi^{c} \tilde \Phi^{d} \tilde \Phi^{e}
\epsilon_{abcde}
+{1\over 2}\partial_t\tilde\Phi^i\partial_t\tilde\Phi^j g_{ij} \tilde\Phi^k
\frac{ \partial_k(g_{tt})^{-1}}{f}\right. \nonumber \\
&& \left. +{1\over {2}}\partial_t\tilde\Phi^i\partial_t\tilde\Phi^j\tilde\Phi^k
\frac{\partial_kg_{ij}}{f}+
{1\over 2}[\tilde\Phi,\tilde\Phi^i][\tilde\Phi^j,\tilde\Phi]\tilde\Phi^k
\frac{\partial_k g_{ij}}{f} + \rm{Fermions} + \rm{higher \ terms}
\right\}.
\end{eqnarray}
Here $\epsilon_{abcde} = \pm1, 0$ is the anti-symmetric symbol
in the $x_\mu,x_\perp$ directions.  Note that $\tilde H_0$ and
$\tilde H_1$ are manifestly independent of $f$, $g$, and $\lambda$.
They depend on ${\cal H}_4$ only through the appearance of the metric
$g_{ij}$ in contractions. As in section \ref{QM}, the factors of $g_{ij}$
would disappear if an orthonormal frame were introduced.  The result is
that the dependence on ${\cal H}_4$ through $g_{ij}$ does not affect
scalars such as $R^2$.

We now wish to calculate $\langle R^2 \rangle$
by treating $H_1$ as a perturbation to $H_0$.
Although the D0-brane
bound state  is normally thought of as degenerate, it is sufficient
to use nondegenerate perturbation theory here.  The point here is that the
ground state degeneracy is associated with supersymmetry and that
states in the corresponding multiplet are generated by the action of the
fermionic zero modes associated with the center of mass.   This follows from
the observation that
the ground states of the N D0-brane bound state must be in one-to-one
correspondence with the ground states of a single D0-brane as required
by the interpretation of the N D0-brane bound state as the eleven dimensional
supergraviton with N units of momentum.  Recall that we have
discarded such center of mass degrees of freedom (whether Bosonic or
Fermionic) as they
decouple from the dynamics of interest.  Thus, the ground state
of our $H_0$ is in fact non-degenerate.

As in section \ref{QM},  the perturbation
$H_1$ is odd under spatial inversion while $H_0$ and $R^2$
are even.  As a result, the first order contribution to $\langle R^2 \rangle$
vanishes.
The leading contribution thus comes from the second order term.  We will make
use of the operator $\bar {P}$, which can be expressed as
the following sum over the eigenstates $|k \rangle_0$ of $H_0$:
\be
\bar{P}= \sum_{k\neq 0} \frac{|k\rangle_0{}_0\langle k|}{ E_k } \equiv
(gN)^{-1/3} {\cal H}_4^{1/12} \lambda^{+1/2}\tilde{\bar P}.
\ee
Note that this definition makes $\tilde{\bar P}$ independent of
$g$, $\lambda$, ${\cal H}_4$, and $f$.
Since the ground state energy of $H_0$ vanishes, the second order
contribution may be written

\bea
\langle R^2 \rangle &-& \langle R^2 \rangle_0 =
\langle \,R^2\bar{P}H_1\bar{P}H_1 + H_1\bar{P}R^2\bar{P}H_1 +
H_1\bar{P}H_1\bar{P}R\,\rangle_0, \nonumber \\  \\
=&& [(gN)^{4/3} f^2 \lambda]
\langle \, \left(Tr \tilde \Phi^2\right) \tilde{\bar{P}} \tilde H_1
\tilde{ \bar{P} } \tilde H_1 +
\tilde H_1 \tilde{\bar{P}} \left(Tr \tilde \Phi^2\right)
\tilde{\bar{P}} \tilde H_1 + \tilde H_1\tilde{\bar{P}} \tilde H_1
\tilde{\bar{P}} \left(Tr \tilde \Phi^2\right)
\,\rangle_0. \nonumber
\label{eq:5}
\eea

As a result, the scaling with $g$, $f$, ${\cal H}_4$, and
$\lambda$ is clearly
\begin{equation}
\label{partres}
 \frac{\langle R^2 \rangle - \langle R^2 \rangle_0}{\langle R^2 \rangle_0}
\sim g^{2/3} {\cal H}_4^{-1/6} f^2,
\end{equation}
in accord with (\ref{qmres}).

Based on the analogy with 't Hooft scaling,
one expects that the dependence on $N$ (in the 't Hooft limit) is
also manifest in equation  (\ref{eq:5}) and that (\ref{partres})
scales like $N^{2/3}$, again in agreement with (\ref{qmres}).  Since,
however, this is less than obvious in the present setting, we now
provide a separate argument for the dependence on $N$.

\subsection{Scaling with N}
\label{tHa}

This new argument is based on the usual 't Hooft power counting, though
it will be done in the framework of Hamiltonian perturbation theory and
we will take due care with regard to the infrared behavior.
Nevertheless, we will need to take as input that certain
features of the expectation values can be read off from the (asymptotic)
perturbation series in a straightforward manner.  This assumption is
in direct parallel with assumption made when 't Hooft scaling is applied
to field theories (i.e., in more than 0+1 dimension)
at strong coupling.  These technical details
will be discussed further below.

As in the familiar 't Hooft argument, the perturbation theory organizes
itself nicely when one studies perturbations about the free (i.e., quadratic)
Hamiltonian.  Thus, we will need to treat even the potential terms
present in $H_0$ above as perturbations.  Note that these terms are
not small.  For this reason as well as others associated directly with
infrared effects as discussed below, we will not be able to truncate
the perturbation expansion at any finite order.  Instead, we need to
consider the entire perturbation series.  We will assume that this
series can be summed in some sense and that this sum describes all of the
physics of interest.

For the argument below we introduce $\tilde f = (gN)^{1/3} {\cal H}^{-1/12}
f$ and consider the rescaled Hamiltonian
\bea
\label{tHham}
\tilde H &=& H_{free} + \delta H =  [(gN)^{-1/3} \lambda^{1/2} {\cal H}^{1/12}]
H  \cr
&=& STr\Biggl( \frac{1}{2} \frac{\tilde P^2}{N} + \frac{N}{4}
[\tilde \Phi,\tilde\Phi]^2 + \tilde fN
\Bigl\{\frac{1}{10}\tilde \Phi^{a} \tilde
\Phi^{b} \tilde \Phi^{c} \tilde \Phi^{d} \tilde \Phi^{e}
\epsilon_{abcde}
+{1\over 2}\partial_t\tilde\Phi^i\partial_t\tilde\Phi_j\tilde\Phi^k
\frac{\partial_k(g_{tt})^{-1}}{f} \cr
&+&
{1\over {2}}\partial_t\tilde\Phi^i\partial_t\tilde\Phi^j\tilde\Phi^k
\frac{\partial_kg_{ij}}{f}+
{1\over 2}[\tilde\Phi,\tilde\Phi^i][\tilde\Phi^j,\tilde\Phi]\tilde\Phi^k
\frac{\partial_k g_{ij}}{f}
\Bigr\}
+ \rm{Fermions} \cr &+& \rm{higher \ terms} \Biggr).
\eea
Here $H_{free}$ contains only the
quadratic terms while $\delta H$ contains all others.

The corresponding system clearly depends on $f$, $g$, and
$\lambda$ only
through the combination $\tilde f$.   The dependence on ${\cal H}_4$
is trivially removed as usual by passing to an orthonormal frame.
The higher terms will in fact
be relevant to our argument, though all that is important is that
they may again be written with only a single explicit factor of $N$
by absorbing appropriate factors of $(gN)$ into the coupling constants
as we have done in rewriting $f$ in terms of $\tilde f$.

We wish to show that the dependence
of $\langle Tr \tilde \Phi^2 \rangle$ and similar expectation values also
arises only through $\tilde f$ (and similar rescaled coupling constants).
This will guarantee that (\ref{partres})
scales with $N^{2/3}$ as desired.   Note that $\tilde H$ and $H$ have
the same ground state.  As a result, expectation values like
$\langle \tilde \Phi^2 \rangle$ are identical for the system defined
by $H$ and the one defined by $\tilde H$.

If $H_{free}$ is to have a normalizeable ground state, then the $\Phi^i$
degrees of freedom must be compactified.  The simplest approach is to replace
each component $(\Phi^i)^a_b$ in the matrix $\Phi^i$ with
$L\sin(\phi^{ia}_b/L)$ for
an angular variable $\phi^{ia}_b$
that lives on a circle and takes values in $[-\pi L, \pi L]$.  In the
limit of large $L$ this reproduces the usual $\Phi^i$ fields which are
valued on the real
line\footnote{Actually, it reproduces $2^{n_{boson}}$
decoupled copies of our
system where $n_{boson}= 9(N^2-1)$ is the total number of such bosonic
degrees of freedom.  However, this will not effect the analysis in any way.}.

Having compactified the bosons, we may
expand the expectation value $\langle Tr \tilde \Phi^2 \rangle =
\langle L^2 \sum_{a,b} g_{ij} \sin(\phi^{ia}_b/L)\sin(\phi^{ib}_a/L)
\rangle$ perturbatively in the
interaction $\delta H$.  At order $k$, each term in this expression will
involve the expectation value in the free ground state of
$L^2 \sum_{a,b} \sin(\phi^{ia}_b/L)\sin(\phi^{ib}_a/L)$ times
$k$ factors of $\delta H$ and $P_{free}$ where

\be
P_{free}= \sum_{n\neq 0} \frac{|n\rangle \langle n|}{ E_n }
\ee
is a sum over eigenstates of $H_{free}$ and $n$ is a $9(N^2-1)$ dimensional
vector giving the mode numbers of the associated momentum eigenstates for
each bosonic degree of freedom.
The states $|n \rangle$ are normalized and so have wavefunctions

\begin{equation}
\langle \phi | n \rangle = \frac{1}{(2 \pi L)^{9(N^2-1)} } \exp(i n \phi)
\end{equation}
with associated $H_{free}$ eigenvalues $E_n = \frac{n^2}{NL^2}$.

Since each term contains the same number of factors of $P_{free}$ and
$\delta H$, we may shift a factor of $N$ from one to the other and write
this series in terms of powers of $\frac{1}{N} \delta H$ and
\be
NP_{free}= \frac{(NL)^2}{n^2}  \sum_{n\neq 0} |n\rangle \langle n|.
\ee
The advantage of this organization is that $\frac{1}{N} \delta H$ has
no explicit dependence on $N$ while the explicit dependence of
$NP_{free}$ on $N$ is only through the combination $NL$.

Each
term of order $k$ has $k$ factors of $NP_{free}$ and has therefore
been written as $(NL)^{2k}$ times
a sum over $k$ mode vectors $n_1,...,n_k$. The summand is the
product of $n_1^{-2} n_2^{-2} ...n_{k}^{-2}$ multiplied both by
one matrix element
of the form
\begin{equation}
\langle n_{j_1} |L^2 \sum_{a,b} \sin(\phi^{ia}_b/L)
\sin(\phi^{ib}_a/L)
|n_{j_2} \rangle \end{equation} and by $k$ matrix elements of the form
$ \frac{1}{N} \langle n_{j_1} | \delta H | n_{j_2} \rangle$,
so that there are $k+1$ matrix elements in all.

Each of the associated operators ($Tr \tilde \Phi^2$ and $\delta H$)
can be written as a sum of monomials in Fermions
and $L \sin(\phi^{ia}_b/L)$.
It is simplest to first run through
the argument ignoring the Fermion terms, and then to address their
effects separately in section A.3 below.
Consider then a purely bosonic matrix element from the above
product having $\alpha$ factors of
$L \sin(\phi^{ia}_b/L)$.  This matrix element takes the form

\begin{equation}
\label{me}
\sum_{a_1,...,a_\alpha}  L^\alpha \langle n_{j_1} |
\sin(\phi^{i_1a_1}_{a_2}/L)\sin(\phi^{i_2a_2}_{a_3}/L)...
\sin(\phi^{i_\alpha a_\alpha}_{a_1}/L) | n_{j_2} \rangle A_{i_1,...,i_\alpha},
\end{equation}
where $A_{i_1,...,i_\alpha}$ is a tensor that does not depend on $L$ or
$N$.

Let us begin by considering `non-diagonal tensors'
$A_{i_1,...,i_\alpha}$ which vanish when any two indices coincide.
The general tensor
$A_{i_1,...,i_\alpha}$ can be written as a sum of such non-diagonal tensors
and `purely diagonal' tensors
(which vanish {\it unless} two or more indices coincide).   It is useful
to note that the states $| n \rangle$ can be written
as the tensor product $\otimes_{i,a,b}
|(n_j)^{ia}_{b} \rangle$, where each momentum eigenstate
$|(n_j)^{ia}_{b} \rangle$ lives in a Hilbert space
$L^2({\bf R})$ associated with the single bosonic degree of freedom
$\phi^{ia}_b$.  Using this observation and again assuming
a non-diagonal tensor
$A_{i_1,...,i_\alpha}$, the matrix elements (\ref{me}) factor as
\begin{eqnarray}
\label{factored}
\sum_{a_1,...,a_\alpha}  \bar \delta_{n_{j_1},n_{j_2}}
L^\alpha \langle (n_{j_1})^{ia_1}_{a_2} &|
\sin(\phi^{i_1a_1}_{a_2}/L)|&(n_{j_2})^{ia_1}_{a_2} \rangle \nonumber \\
 &&\times\langle (n_{j_1})^{ia_2}_{a_3}|\sin(\phi^{i_2a_2}_{a_3}/L)|(n_{j_2})^{ia_1}_{a_2}
 \rangle...\\
 &&\times \langle (n_{j_1})^{ia_\alpha}_{a_1} |
\sin(\phi^{i_\alpha a_\alpha}_{a_1}/L) | (n_{j_2})^{ia_\alpha}_{a_1}
\rangle A_{i_1,...,i_\alpha}, \nonumber
\end{eqnarray}
where
$\bar \delta_{n_{j_1},n_{j_2}}$ is a Kronecker delta function
that sets $n_{j_1}$ equal to $n_{j_2}$ except for those components of
$n_{j_1}, n_{j_2}$ that still appear explicitly in (\ref{factored}).

Note that each factor
\begin{equation}
\label{Kds}
 \langle (n_{j_1})^{ja}_{b} |
\sin(\phi^{ja}_{b}/L)|(n_{j_2})^{ja}_{b} \rangle = \frac{1}{2i} \left(
\delta_{(n_{j_1})^{ja}_b, (n_{j_2})^{ja}_b + 1} -
\delta_{(n_{j_1})^{ja}_b, (n_{j_2})^{ja}_b - 1} \right)
\end{equation}
is independent of $N$ or $L$ while the
sum over $a_1,...,a_\alpha$ contributes $N^\alpha$
similar terms.  The sum over spatial directions
$i_1,...i_\alpha$ generates a purely
numerical term controlled by the tensor
$A_{i_1,...,i_\alpha}$ which is thus independent of $N$ and $L$.

Consider now the full contribution from a term at order $k$ which is
built entirely from non-diagonal tensors.  This yields
a factor of $(NL)^{2k + \alpha_1 + \alpha_2 + ... + \alpha_k + \alpha_{k+1}}$
times a sum over $n_1,...,n_k$.  This final sum contains no explicit
factors of $N$ or $L$.  Recall that components of $n_{j_1}, n_{j_2}$
that do not appear
explicitly in (\ref{factored}) are controlled by the Kronecker delta
function
$\bar \delta_{n_{j_1},n_{j_2}}$, while those appearing explicitly
in (\ref{factored}) are controlled by the Kronecker delta functions in
(\ref{Kds}).  Thus, the final sum
contains at least one Kronecker delta function
for each component of each $n_i$ over which the sum is performed.
As a result, the number of terms
contributing to the final sum is independent of $N$ or $L$ (though it does
depend on $\alpha_1,\alpha_2,...\alpha_{k+1}$).  We conclude that
the full contribution from each term at order $k$ is a purely numerical
factor times
$(NL)^{2k + \alpha_1 + \alpha_2 + ... + \alpha_k + \alpha_{k+1}}$.
The important observation here is that the result depends only
on the product $NL$.

Let us now consider a term with a purely diagonal tensor.  While the bosonic
fields $\phi^{ia}_b$
 involved in a given matrix element might still be distinct, they
need not necessarily be so.  Thus there
may be factors of $\langle (n_{j_1})^{ia}_{b} |
L^m \sin^m(\phi^{ia}_{b}/L)|(n_{j_2})^{ia}_{b} \rangle$
for $m > 1$.  Such matrix elements are again of order one times Kronecker
deltas.  However, there are only $N$ such terms compared to the $\left(
{N \atop m} \right)$ terms involving $m$ different factors of the form
$\langle (n_{j_1})^{ia}_{b} | L
\sin(\phi^{ia}_{b}/L)|(n_{j_2})^{ia}_{b} \rangle$.  Thus, at a given
order in $L$, the contribution from terms with repeated bosonic fields
in a given matrix element is suppressed by factors of $N$ relative
to terms with distinct bosonic fields in each matrix element.
It follows that the contribution
of purely diagonal tensors is again of the form
$(NL)^{2k + \alpha_1 + \alpha_2 + ... + \alpha_k + \alpha_{k+1}}$ times
a numerical factor plus sub-leading terms in $N$.

We should now take the large $L$ limit and then take the limit of large $N$.
Clearly, however, this limit fails to converge if we truncate the expansion
at a fixed order $k$.  From the field theory perspective, this is the usual
infrared divergence of perturbation theory in low dimensional systems.
Thus, we should think of using the perturbation series as a whole to extract
information about the full function $\langle Tr \tilde \Phi^2 \rangle$ of
$N,L$.  That is, the series must first be summed and then the large $L$ limit
may be taken.  We have seen that the sum has one contribution from
terms that involve only the product $NL$ and another from terms
that are sub-leading in $N$ (at a fixed order in $L$).  We now make the
(reasonable but not necessarily justified) assumption that at large $N$
we may neglect the terms that are sub-leading in $N$ at a given order in
$L$.    Note that this assumption mirrors that of the field theory
application of 't Hooft scaling at strong coupling.  There, the 't Hooft
scaling argument assumes that terms sub-leading in $N$ at a given order
in the 't Hooft coupling can be neglected even though one is not
in the perturbative regime.

After discarding such sub-leading terms,
What remains is a function only of the product $NL$.  Thus,
we expect that at large $N$ the function
$\langle Tr \tilde \Phi^2 \rangle$ depends only
on the product $NL$.  Finally, the full
Hamiltonian
should have
a normalizeable ground state\footnote{This is
the assumption that the presence of
a distant D4-brane does not completely disrupt the D0 bound state and is
to be expected on the basis of both supersymmetry arguments and supergravity
physics.
Note however that this assumption would be false if we applied it only
to the terms explicitly displayed in (\ref{tHham}) since the
quintic term is unbounded below.}  even for infinite $L$ so that the
limit $L \rightarrow \infty$ of
$\langle Tr \tilde \Phi^2 \rangle$ is finite and thus independent of $L$.
Since, however, this function depended on $N$ only through the product $NL$,
it follows immediately that the large $L$ limit of
$\langle Tr \tilde \Phi^2 \rangle$ is also independent of $N$.
This is exactly the result we desired to show.

\subsection{Fermions}

Of course, we must still consider the effect of the Fermion terms.
Note that there is an important effect of the Fermions already at
the kinematic level, in that the Fermion zero modes cause the free
ground state to be massively degenerate.  For our free Hamiltonian, all
Fermion degrees of freedom contribute to such a degeneracy, not just
those associated with the center of mass.  We expect this degeneracy
to be completely lifted, but we should ensure that we are indeed
computing the expectation value in the correct ground state.

As is usual in degenerate perturbation theory, we use a basis of free
ground states in which the perturbation $\delta H$ is diagonal.
This ensures that the perturbation series takes the same form as in
(\ref{tHa}), and that
there is no danger of dividing by a vanishing energy denominator $E_n$.
Since this degeneracy should
be completely lifted by the interaction, one particular state
in this basis will lead to the true ground state.  Let us call this
state $|0\rangle.$  In contrast, we will denote the original
Fermion vacuum $|vac \rangle.$
Since $\delta H$ is invariant under both SO(10) rotations and SU(N), both
the true ground state and $|0\rangle$ must be as well.

Now, recall that any operator in the space of free ground states
can be expressed as a finite polynomial in $\tilde \Theta^{ia}_b.$
Consider in particular this representation of the projection operator
$|0\rangle \langle 0 |$.  Invariance under SU(N) tells us that this
operator may be expressed as a sum of the traces $Tr (\tilde \Theta^{1_i}
... \tilde \Theta^{i_k} )$.  Thus, we may write

\begin{equation}
\label{vactrans}
|0\rangle = \sum_\beta A^\beta_{i_1,...,i_\beta} \tilde \Theta^{i_1}...
\tilde \Theta^{i_\beta} |vac \rangle,
\end{equation}
for an appropriate collection of tensors
$A^\beta_{i_1,...,i_\beta} \tilde \Theta^{i_1}$.  These tensors are
clearly independent of $L$.  To investigate their dependence on $N$,
it is important to understand the normalization of $\tilde \Theta^i$
properly.  From \cite{tvr1}, we find (in terms of our conventions)
that the original action $S$ contains a factor of
$\frac{{\cal H}_4^{1/4} \lambda^{3/2}}{g}$  in front of the Fermion
Kinetic term.  Thus, the anticommutation relations of the $\Theta^i$
Fermions take the form $\{\Theta^{ia}_b ,\Theta^{ia}_b \}_+  \sim g
{\cal H}_4^{-1/4} \lambda^{-3/2}$.
However, the rescaled fields are $\tilde \Theta^i = \lambda^{3/4}
{\cal H}_4^{1/8} (gN)^{-1/2} \Theta^i$,
so we have $\{\tilde \Theta^{ia}_b,
\tilde \Theta^{ia}_b \}  \sim \frac{1}{N}$.  Normalization of
$|0\rangle$ and the property that
$\langle vac | \Theta^{i_1a_1}_{b_1}...\Theta^{i_ka_k}_{b_k} | vac \rangle$
vanishes unless the Fermions
$\Theta^{i_1a_1}_{b_1}...\Theta^{i_ka_k}_{b_k}$ occur in pairs
then tells us that the tensors
$A^\beta_{i_1,...,i_\beta}$ are independent of $N$ to leading order in $N$.

Let us now briefly review the results of the purely bosonic argument above.
At order $k$ in perturbation theory, we find $2k$ factors of $NL$ from
the explicit factors of $N$ in $\delta H$ and from the energy denominators
in $P_{free}.$  In addition, each boson contributes at most one
factor of $L$ and
a factor of $N$ from summing over an SU(N) index.

Adding Fermions introduces factors of $\tilde \Theta^i$ into the matrix
elements.  Each $\tilde \Theta^i$ introduces an additional sum over $N$, just
as occurred for each boson.  However, factors of $\tilde \Theta^i$ are
not associated with factors of $L$ since the infrared regularization affects
only the bosonic zero modes.  Consider a given term that is a
product of matrix elements of the form (\ref{me}), perhaps with additional
Fermion factors.  Of course, two of the mode states appearing in each
factor must in fact be the vacuum state $|0\rangle.$

We may write
the $|0\rangle$ states in terms of $|vac \rangle$ by using (\ref{vactrans}).
The fact that the free system is a direct product of Boson
and Fermion degrees of freedom may be used to write each product of matrix
elements in terms of two factors.  One of these factors involves
only the bosonic degrees of freedom and is much like (\ref{me}) above.
The other is just the expectation value of
a product of Fermion operators $\tilde \Theta^{ia_1}_{a_2}$ in the state $|vac
\rangle.$  The relevant point is that such expectation values vanish
unless each $\tilde \Theta^{ia_1}_{a_2}$ that occurs in the product
matches with some other Fermion operator $\tilde \Theta^{jb_1}_{b_2}$; that is,
the Fermion operators must
combine in pairs and create a factor setting $(a_1,a_2)=(b_1,b_2)$.

This introduces two constraints which may remove up to two sums
over SU(N) indices.  The first time a Fermion is paired within a single
trace, two new constraints are in fact imposed.
Typically, however, one of the SU(N) indices
on the   $\tilde \Theta^{ia_1}_{a_2}$ factor will already have been fixed
by a previous constraint.  Therefore, in general
we should only count
this constraint as removing a single factor of $N$ from our
counting.  Thus, these constraints remove one factor of $N$ for each pair
of Fermions\footnote{
If a trace involves
only Fermions (and no bosons; i.e., the sort of term
that appears in (\ref{vactrans})), then when the last Fermion from a given
trace is paired it may be that all of
the constraints are already in place.  In this case, no new constraints
are introduced.  Since, however, this can occur no more than once in
any trace, and since our scheme undercounted the constraints imposed
when the first Fermion from the trace was paired, it is safe to state
that the number of constraints imposed is at least the number of pairs
created.}.

If in addition we recall that with the current normalizations
we have $\{ \tilde \Theta^i, \tilde \Theta^j \}_+ \sim \frac{1}{N}$, then
we see that the creation of each pair removes at least one sum over
SU(N) indices and
provides an additional factor of $N^{-1}$.  Recall that there was originally
one sum over SU(N) indices (and thus one potential factor of $N$)
for each Fermion and thus two factors for each pair.
We therefore see that to leading order the Fermions
contribute $N^0 \sim 1$ and $L^0 \sim 1$.
As a result, the leading factors of $N$ continue
to appear precisely paired with the factors of $L$ and we may again argue
that the result is independent of $N$ in the limit of large $L$.

%%%%%%%%%%%%%%%%%%%%%%%%%%%%%%%%


\begin{thebibliography}{99}
\bibitem{bfss}  T. Banks, W. Fischler, S.H. Shenker and L. Susskind, "M Theory As A Matrix Model: A Conjecture
                ", Phys.Rev. D55 (1997) 5112-5128, {\tt hep-th/9610043}.
\bibitem{malda} J. Maldacena, ``The large N limit of superconformal field
theories and supergravity,'' {\it Adv. Theor. Math. Phys.} {\bf 2} (1998)
231, hep-th/9711200.
\bibitem{imsy}  N. Itzhaki, J. M. Maldacena, J. Sonnenschein, S. Yankielowicz,"Supergravity and The Large N Limit of Theories With Sixteen Supercharges
                ", Phys.Rev. D58 (1998) 046004, {\tt hep-th/9802042}.
\bibitem{KL} D. Kabat and D. Lowe, D.~Kabat, G.~Lifschytz and D.~A.~Lowe,
``Black hole entropy from non-perturbative gauge theory,''
hep-th/0105171;
%%CITATION = HEP-TH 0105171;%%
D.~Kabat, G.~Lifschytz and D.~A.~Lowe,
``Black hole thermodynamics from calculations in strongly-coupled gauge  theory,''
Phys.\ Rev.\ Lett.\  {\bf 86}, 1426 (2001)
[Int.\ J.\ Mod.\ Phys.\ A {\bf 16}, 856 (2001)]
[hep-th/0007051].
%%CITATION = HEP-TH 0007051;%%

\bibitem{Dft} A.~Giveon and D.~Kutasov,
``Brane dynamics and gauge theory,''
Rev.\ Mod.\ Phys.\  {\bf 71}, 983 (1999)
[hep-th/9802067].
%%CITATION = HEP-TH 9802067;%%
\bibitem{SV}  A.~Strominger and C.~Vafa,"Microscopic Origin of the Bekenstein-Hawking Entropy",
              Phys.\ Lett.\  {\bf B379} 99 (1996), {\tt hep-th/9601029}.
\bibitem{tvr2}  W. Taylor, M. Van Raamsdonk, "Multiple Dp-branes in Weak Background Fields
                ", Nucl.Phys. B573 (2000) 703-734, {\tt hep-th/9910052}.
\bibitem{mye1}  R.C. Myers, "Dielectric-Branes",JHEP 9912 (1999) 022, {\tt hep-th/9910053}.
\bibitem{pol1}  J. Polchinski, "M-Theory and the Light Cone
                ", Prog.Theor.Phys.Suppl. 134 (1999) 158-170, {\tt hep-th/9903165}.
\bibitem{tse1}  A.A. Tseytlin, "Born-Infeld action, supersymmetry and string theory
                ", {\tt hep-th/9908105}.
\bibitem{bai1}  P. Bain, "On the non-abelian Born-Infeld action", {\tt hep-th/9909154}.
\bibitem{tvr1}  W. Taylor, M. Van Raamsdonk, "Multiple D0-branes in Weakly Curved Backgrounds
                ", Nucl.Phys. B558 (1999) 63-95, {\tt hep-th/9904095}.
\bibitem{ch}    M Clauson and B. Halpern, "Supersymmetric Ground
                State Wave Functions", Nucl. Phys. B250 (1985)
                689.
\bibitem{indios} S.~P.~Trivedi, S.~Vaidya, "Fuzzy Cosets and their Gravity Duals
                ", JHEP 0009 (2000) 041, {\tt hep-th/0007011}.
\bibitem{gjs}   C. Gomez, B. Janssen, P. J. Silva, "Dielectric branes and spontaneous symmetry breaking
                ", {\tt hep-th/0011242}
\bibitem{clt}   J. Castelino, S. Lee, W. Taylor, "Longitudinal 5-branes as 4-spheres in Matrix theory
                ", Nucl.Phys. B526 (1998) 334-350, {\tt hep-th/9712105}.
\bibitem{thooft} G. t'Hooft, " A Planar Diagram Theory of Strong Interactions", Nucl. Phys. B72 (1974) 461.
\bibitem{SM}    S.~Surya and D.~Marolf, ``Localized branes and black holes,''
                Phys.\ Rev.\ D {\bf 58}, 124013 (1998), {\tt hep-th/9805121}.
\bibitem{mar1}  D. Marolf, A. W. Peet,
"Brane Baldness vs. Superselection Sectors
                ",Phys.Rev. D60 (1999) 105007, {\tt hep-th/9903213}.
\bibitem{HM}    G.~T.~Horowitz and D.~Marolf, ``Where is the information stored in black holes?,''
                Phys.\ Rev.\ D {\bf 55}, 3654 (1997), {\tt hep-th/9610171}.
                %%CITATION = HEP-TH 9610171;%%
\bibitem{Tseytlin} A.~A.~Tseytlin,%``Composite BPS configurations of p-branes in 10 and 11 dimensions,''
                   Class.\ Quant.\ Grav.\ {\bf 14}, 2085 (1997),{\tt hep-th/9702163}.
                %%CITATION = HEP-TH 9702163;%%
\bibitem{BREJS} E.~Bergshoeff, M.~de Roo, E.~Eyras, B.~Janssen and J.~P.~van der Schaar,
                %``Intersections involving monopoles and waves in eleven dimensions,''
                Class.\ Quant.\ Grav.\ {\bf 14}, 2757 (1997), {\tt hep-th/9704120}.
                %%CITATION = HEP-TH 9704120;%%
\bibitem{AIRV} I.~Y.~Aref'eva, M.~G.~Ivanov, O.~A.~Rytchkov and I.~V.~Volovich,
               %``Non-extremal localized branes and vacuum solutions in M-theory,''
               Class.\ Quant.\ Grav.\ {\bf 15}, 2923 (1998), {\tt hep-th/9802163}.
               %%CITATION = HEP-TH 9802163;%

\end{thebibliography}
\end{document}